\documentclass[fleqn,usenatbib]{mnras}
\usepackage{newtxtext,newtxmath}
\usepackage{comment}
\usepackage{multirow}
\usepackage{mathtools}
\usepackage[T1]{fontenc}
\usepackage{ae,aecompl}

\usepackage{graphicx}	
\usepackage{amsmath}	
\usepackage[ruled,vlined]{algorithm2e}
\usepackage{algpseudocode}
\usepackage{enumitem} 
\usepackage{xfrac} 

\setlist[itemize]{leftmargin=*}

\newcommand{\Lsun}{L_\odot}

\newcommand{\zsol}[1]{{#1}\:\mathrm{Z_\odot}}
\newcommand{\msol}[1]{{#1}\:\mathrm{M_\odot}}
\newcommand{\Sigmasol}[1]{{#1}\:\mathrm{M_\odot\:pc^{-2}}}
\newcommand{\rhosol}[1]{{#1}\:\mathrm{M_\odot\:pc^{-3}}}

\newcommand{\ratesol}[1]{{#1}\:\mathrm{M_\odot yr^{-1}}}
\newcommand{\collratesol}[1]{{#1}\:\mathrm{yr^{-1}}}

\newcommand{\mstar}{\texttt{MSTAR}}
\newcommand{\bifrost}{\texttt{BIFROST}}
\newcommand{\sevn}{\texttt{SEVN}}

\title[SMSs with stellar BH cores in faint LRDs]{{Supermassive stars with embedded stellar black hole cores: dense assembling star clusters as faint multiple Little Red Dot systems}}
\author[A. Rantala]{Antti Rantala$^{1,2,3}$\thanks{E-mail: antti.rantala@ast.cam.ac.uk}\\
$^{1}$Institute of Astronomy, University of Cambridge, Madingley Road, Cambridge CB3 0HA, UK\\
$^{2}$Kavli Institute for Cosmology, Cambridge (KICC), University of Cambridge, Madingley Road, Cambridge CB3 0HA, UK\\
$^{3}$Max-Planck-Institut f\"ur Astrophysik, Karl-Schwarzschild-Str. 1, D-85748, Garching, Germany\\
}
\date{Accepted XXX. Received YYY; in original form ZZZ}
\pubyear{2026}

\begin{document}
\label{firstpage}
\pagerange{\pageref{firstpage}--\pageref{lastpage}}
\maketitle

\begin{abstract}
Numerical simulations have established that star clusters with densities comparable to the high redshift ($z>6$--$10$) James Webb Space Telescope (\textit{JWST}) proto globular clusters may build up extremely massive (EMSs; $m_{\star}>\msol{1000}$) or even supermassive stars (SMSs; $m_{\star}>\msol{10000}$) and potentially intermediate mass black holes (IMBHs) through runaway stellar collisions. Using direct simulations of assembling star clusters including post-Newtonian black hole dynamics and stellar evolution, we demonstrate that in such dense environments ($\Sigma_\mathrm{h} \gtrsim \Sigmasol{10^6}$) stellar BHs ($m_\bullet \lesssim \msol{60}$), driven by rapid mass segregation and relaxation effects within the sphere of influence of the EMSs/SMSs, may strongly interact with the extremely massive stars and become embedded within their gaseous layers. We suggest that this quasi-star (QS) like embedded BH phase is a natural outcome of the runaway formation of EMSs/SMSs in the densest star clusters. The QS phase may be $1$--$2$ orders of magnitude longer in duration than the lifetime of the SMS, enabling an extended growth period by stellar collisions, and allows the formation of embedded gravitational wave sources if the QS captures more than a single stellar BH. The star cluster assembly region sizes ($\sim100$ pc), QS masses ($\gtrsim \msol{10^4}$) and their proximity to young, massive blue star forming clumps are consistent with the faint population of multiple little red dots (LRDs) recently discovered by the \textit{JWST}.
\end{abstract}



\section{Introduction}

Astrophysical models in which a compact object such as a neutron star (NS) or a black hole (BH) is embedded within the gaseous layers of a star have for a long time been constructed out of theoretical curiosity and invoked to interpret various unexplained observational results (e.g. \citealt{Clayton1975, Oncins2022, Bellinger2023}) especially in the context of binary stellar evolution (e.g. \citealt{Sparks1974, Paczynski1976, MacLeod2015, CruzOsorio2020}) and supermassive black hole (SMBH) seed formation. Such more or less exotic astrophysical objects include red giant stars with NS cores (Thorne-\.{Z}ytkow objects, \citealt{ThorneZytkow1977, Cannon1992, Hutilukejiang2018, Farmer2023, Everson2024, Nathaniel2025, Williams2025}) or quasi-stars (QSs; \citealt{Begelman2008,Volonteri2010,Ball2011,Schleicher2013,Coughlin2024}) in hydrostatic equilibrium with their central BH cores. Quasi-stars have been considered as relatively short-lived black hole gas envelope phases of the low metallicity direct collapse gas cloud scenario channel (DCBH; \citealt{Bromm2003, Begelman2006, Dijkstra2008, Begelman2010, Agarwal2012, Latif2013, Habouzit2016, Regan2017, Wise2019, Chon2021, Jeon2025}) for SMBH seed formation \citep{Rees1978, Volonteri2008, Natarajan2014, Inayoshi2020}. Recently, attention for such embedded BH scenarios has been reinvigorated by the discovery of a novel and enigmatic class of objects now known as little red dots (LRDs; \citealt{Matthee2024}) by the James Webb Space Telescope (\textit{JWST}). LRDs are characterized, among other properties, by their abundance at $z\gtrsim4$ \citep{Kokorev2024,Kocevski2025}, compact ($\lesssim 100$ pc) sizes (e.g. \citealt{Labbe2025}), and peculiar spectral energy distributions (SEDs; \citealt{Barro2024,Harikane2023b,Williams2024,Yue2024,DEugenio2026,Inayoshi2026}) including broad emission lines, X-ray weakness and weak variability, V-shaped spectral energy distributions, and strong Balmer absorption features. Despite their compactness, a large fraction of LRDs show evidence of multiple associated components or exhibit highly asymmetric structures despite appearing as single sources \citep{Rinaldi2025}. Recent studies have also indicated small scale sub-kpc clustering of the LRDs \citep{Tanaka2024, Merida2025}, and individual faint LRD pairs have been detected down to separations of $\sim70$ pc \citep{Yanagisawa2026}. In addition, LRDs are commonly associated with blue star forming companions with a few hundred pc to kpc offsets \citep{Merida2025, Golubchik2025, Baggen2025, Chen2025}, while accretion powered sources known as Little Blue Dots \citep{Brazzini2026} differ from LRDs in their appearance likely by their amount of gas and its geometry around the SMBH. A fraction of LRDs may host SMBHs considerably overmassive to their host galaxies or peculiar extremely low metallicities (e.g. \citealt{Juodzbalis2025, Maiolino2025}) that may pose a challenge to the theoretical SMBH seeding scenarios \citep{Dayal2026}.

While the nature of the LRDs remains under active debate, several theoretical scenarios involving massive BHs or their progenitors have been proposed to interpret their remarkable properties \citep{Inayoshi2025b}. Number density and spectral signature arguments have been invoked to explain LRDs as accreting SMBHs embedded in dense gaseous atmospheres or broad line regions \citep{Inayoshi2025, Kido2025, Madau2025, Maiolino2025b, Sneppen2026}, or so-called black hole stars \citep{Naidu2025b, deGraaff2025}, DCBHs \citep{Cenci2025, Jeon2026, Pacucci2026}, late-stage quasi-stars \citep{Begelman2026,Santarelli2026}, supermassive stars (SMSs; \citealt{ Martins2020, Pacucci2025, Nandal2026}) together with their host star clusters \citep{Chisholm2026, Williams2026}, expelled gas cocoon shells \citep{Nandal2026b}, accretion disks \citep{Zwick2025}, or tidal disruption events of stars by the SMBH seeds in collapsing star clusters \citep{Bellovary2025} combined with rapid gas accretion \citep{Kritos2025}. The various proposed models are not necessarily mutually exclusive: very recently, \cite{RomanGarza2026b} proposed an accreting QS model that unifies and generalises the SMS and QS scenarios over a broad metallicity range. Thus far, most of the QS models involving embedded BHs are founded on the DCBH scenario, or gas accreting proto star like SMSs \citep{Hosokawa2012, Latif2016, RamirezGaleano2025} that form BH cores via general relativistic instability \citep{Chandrasekhar1964, Haemmerle2021, RomanGarza2026}. In this study we present a channel for the formation of such embedded QS like objects from another widely studied SMBH seed formation channel: runaway stellar collisions.

Despite various physical and numerical modelling uncertainties, it is now generally agreed that runaway growth by stellar collisions \citep{Schneider2025} may occur in dense star clusters (e.g. \citealt{PortegiesZwart1999,PortegiesZwart2004,Freitag2006a,Mapelli2016,ArcaSedda-DRAGON2a,GonzalezPrieto2024,Rantala2024b,Vergara2025}). In general, \textit{JWST} $z\gtrsim6$--$10$ proto globular cluster like environments (e.g. \citealt{Adamo2024, Mowla2024, Bradac2025, Abdurrouf2025}) are thought to be sufficiently dense to reach the SMS regime via collisional runaways \citep{Rantala2025b, Vergara2025b}. The runaway growth may result in the formation of an intermediate mass black hole (IMBH), unless catastrophic mass loss events from a subset of the stellar collisions \citep{RamirezGaleano2025,RomanGarza2026} or exceptionally high wind mass loss rates (e.g. \citealt{Belkus2007, Glebbeek2009, Torniamenti2025}) can disrupt the collisionally growing stars. The lifetime of such collisionally grown very massive (VMS; $>\msol{100}$), extremely massive (EMS; $>\msol{1000}$) or SMSs ($>\msol{10000}$; \citealt{Hoyle1963, Fowler1966, Shapiro1979, Baumgarte1999, Denissenkov2014, Gieles2018,Charbonnel2023,Gieles2025}) is expected to be short, less than a few Myr (e.g. \citealt{Nandal2025}). However, due to the process of rejuvenation by collisions \citep{Leonard1989,Lombardi2002,Hurley2002} or gas accretion \citep{Nandal2025}, the massive stellar collision product may have a longer lifetime compared to typical massive ($\lesssim \msol{150}$) stars that do not experience collisions. Dynamical friction \citep{Chandrasekhar1943} and the mass segregation \citep{Spitzer1971} process ensure that stellar BHs and their progenitor stars efficiently sink into the centre of the dense cluster, leading to the build-up of up to tens of stellar BH within the cluster core. Within a brief time window of $1$--$2$ Myr after the formation of the first stellar BHs but before the EMS reaches the end of its life and forms an IMBH, the stellar BHs and the EMSs/SMSs may strongly interact. We propose that instead of the disruption of the star, at least a subset of such interactions may lead to the formation of a QS like object.

In the previous numerical work, collisions between black holes and stars whose masses exceed those of the black holes have been regarded as a somewhat numerical peculiarity (e.g. \citealt{Rizzuto2021, Rantala2025b, Rantala2026, Vergara2025, Mestichelli2026}) highlighting the obvious limitations of the simplified sub-resolution tidal disruption event prescriptions of the simulation codes (e.g. \citealt{Kochanek1992, Rizzuto2023}). Instead, we argue that such interactions, despite the current modelling limitations, are a plausible and frequent outcome of models that build up the EMSs/SMSs through repeated stellar collisions in dense star clusters. Such a quasi-star like configuration is likely to affect the further evolution of the stars: compared to SMSs, QSs have considerably extended lifetimes (e.g. \citealt{Begelman2026, RomanGarza2026b, Santarelli2026}). This extended evolutionary phase may enable prolonged mass growth via collisions or gas accretion, however, the chemical element production \citep{RomanGarza2026b} within the star likely gets suppressed. Finally, we speculate that QS like SMSs that acquire multiple stellar BHs may constitute a class of exotic gas embedded gravitational wave (GW) sources in assembling star clusters. 

This study is structured as follows. In Section \ref{section: 2} we briefly review our simulation code \bifrost{} and the initial star cluster models. We detail our results concerning BH EMS/SMS interactions in Section \ref{section: 3}, and interpret our findings in terms of early LRD evolution in Section \ref{section: 4}. Finally, we summarise and conclude in Section \ref{section: 5}.

\section{Methods and star cluster models}\label{section: 2}

\subsection{Simulation code}
The simulations for this study first described in \cite{Rantala2026} were performed using the GPU accelerated direct summation \textit{N}-body code \bifrost{} \citep{Rantala2023} based on the hierarchical fourth order forward integrator \citep{Rantala2021} and a selection of secular and regularised few-body solvers (e.g. \citealt{Rantala2020, Wang2020, Rantala2024b}) with post-Newtonian equations of motion up to the order PN3.5 for BHs. For stellar evolution, \bifrost{} is coupled to the fast stellar population synthesis code \sevn{} \citep{Iorio2023,Mapelli2020,Spera2015,Spera2017} based on PARSEC stellar tracks \citep{Bressan2012,Chen2015,Nguyen2022,Costa2025} up to $\msol{600}$. Stars can collide and merge if their radii overlap, losing up to $7.5\%$ of their mass per collision using a mass ratio based fitting formula (e.g. \citealt{Glebbeek2013}). The collision product age is set as in \cite{Mapelli2016} while taking the mass loss into account \citep{Rantala2026}. This rejuvenation procedure avoids the excess rejuvenation issues reported by \cite{Vergara2025b} as the models result in zero rejuvenation when the mass of the secondary star is negligible ($q=m_\mathrm{2}/m_\mathrm{1} \rightarrow 0$). The stellar lifetimes are extrapolated from the highest mass PARSEC tracks available, and the combined rejuvenation + lifetime assignment procedure leads to maximum EMS/SMS lifetimes comparable to other $\textit{N}$-body studies (e.g. \citealt{Vergara2025}). We emphasise that such $\textit{N}$-body rejuvenation recipes \citep{Hurley2002} treat the stellar rejuvenation in a simplified manner, as in more detailed models the rejuvenation of the massive star depends on the mixing and convection treatment \citep{Gaburov2008b, Ballone2023, RamirezGaleano2025} as well as the merger and gas accretion history of the star. Stars more massive than $\msol{600}$ lose mass by metallicity dependent \cite{Vink2018} winds (referred to as \textit{stronger} winds) with an option to impose a \cite{Humphreys1994} like wind limit to mimic the \textit{level C} stellar evolution prescriptions of \texttt{NBODY6++GPU} \citep{Kamlah2022,Vergara2025} (\textit{weaker} winds). Stars above the pulsational pair instability mass gap collapse at the end of their lives forming IMBHs that may further grow by merging with stellar BHs and in hierarchical setups with other IMBHs, or by tidally disrupting stars. In \cite{Rantala2026}, the disruption separation between a star and a BH is $R_\mathrm{disrupt} = \max{(R_\mathrm{\star}, 1.3 (m_\bullet/m_{\star})^{1/3}R_\mathrm{\star})}$ in which $m_\bullet$ and $m_{\star}$ are the BH and stellar masses, respectively \citep{Kochanek1992}. In TDEs, $50\%$ of stellar material is accreted onto the BH. In this study, we now re-consider the type of interaction for BHs and stars with $m_\star \gg m_\bullet$. However, as we do not perform new simulations in this study, we leave the QS like SMS model implementation into \bifrost{} as future work.

\begin{table*}
    \centering
    \begin{tabular}{lccccccccc}
    \hline
        model(s) & $N_\mathrm{density}$ & $N_\mathrm{random}$ & model type & $Z/\zsol{}$ & winds & $r_\mathrm{h}$ [pc] & $\rho_\mathrm{h}$ $[\rhosol{}]$ & $\Sigma_\mathrm{h}$ $[\Sigmasol{}]$\\
        \hline
        HD[1--9]Z2 & 9 & 3 & hierarchical & $0.10$ & Vink 2018 & $0.08$--$2.16$ & $3.0\times10^3$--$6.5\times10^7$ & $5.0\times10^3$--$3.9\times10^6$\\ 
        HD9Z1      & 1 & 3 & hierarchical & $0.01$ & Vink 2018 & $0.08$--$0.10$ & $3.0\times10^7$--$6.5\times10^7$ & $2.4\times10^6$--$3.9\times10^6$\\
        ID9Z1      & 1 & 3 & isolated     & $0.01$ & weaker than Vink 2018 & $0.08$--$0.10$ & $3.0\times10^7$--$6.5\times10^7$ & $2.4\times10^6$--$3.9\times10^6$\\
         \hline
    \end{tabular}
    \caption{The simulation sample of this study. The simulations are a sub-sample of the large collection of models originally described in \citet{Rantala2026}. Our main focus are the densest models HD9Z1, HD9Z2 and ID9Z1 that efficiently form EMSs and even SMSs (in the case of weaker winds).}
    \label{table: 1}
\end{table*}

\begin{table}
    \centering
    \begin{tabular}{lccc}
\hline   
Central & min--max $r_\mathrm{h}$ & min--max $\rho_\mathrm{h}$ & min--max $ \Sigma_\mathrm{h}$\\
sub-cluster & pc & [$\rhosol{10^5}$] & [$\Sigmasol{10^5}$]\\
\hline   
HD1 & $1.67$--$2.16$ & $0.03$--$0.06$ & $0.05$--$0.08$ \\ 
HD2 & $1.14$--$1.47$ & $0.09$--$0.20$ & $0.11$--$0.18$ \\ 
HD3 & $0.78$--$1.00$ & $0.30$--$0.64$ & $0.23$--$0.39$ \\ 
HD4 & $0.53$--$0.68$ & $0.94$--$2.02$ & $0.51$--$0.84$ \\ 
HD5 & $0.36$--$0.46$ & $2.97$--$6.41$ & $1.09$--$1.82$ \\ 
HD6 & $0.24$--$0.32$ & $9.42$--$20.33$ & $2.35$--$3.93$ \\ 
HD7 & $0.17$--$0.22$ & $29.87$--$64.45$ & $5.08$--$8.48$ \\ 
HD8 & $0.11$--$0.15$ & $94.68$--$204.32$ & $10.95$--$18.29$ \\ 
HD9 & $0.08$--$0.10$ & $300.14$--$647.69$ & $23.63$--$39.47$ \\ 
\hline   
    \end{tabular}
    \caption{The initial half-mass sizes $r_\mathrm{h}$, as well as 3D and 2D initial half-mass densities $\rho_\mathrm{h}$ and $\Sigma_\mathrm{h}$ of the central clusters of the nine hierarchical models. The isolated model ID9Z1 corresponds to the central cluster of HD9.}
    \label{tab: cluster-densities}
\end{table}

\begin{figure}
    \centering
    \includegraphics[width=0.7\linewidth]{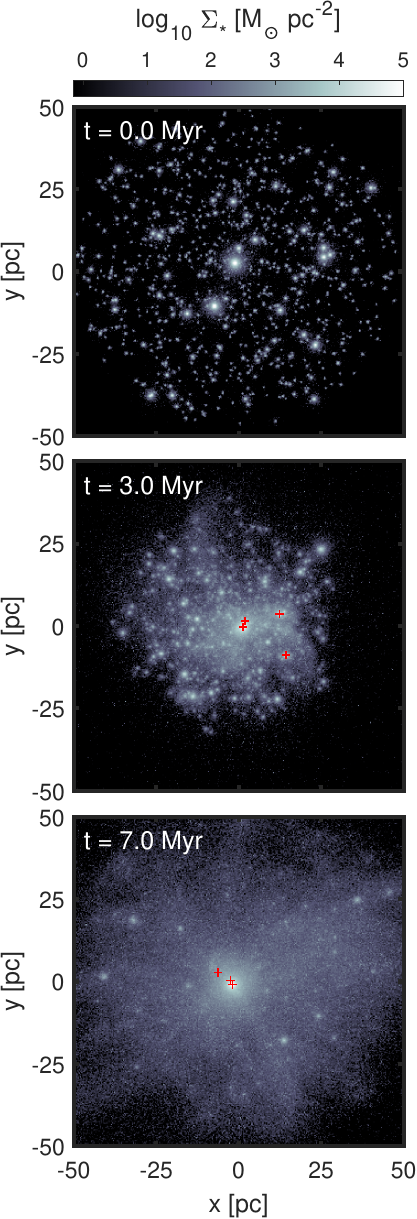}
    \caption{The rapid hierarchical assembly of a massive star cluster from HD5Z2 initial conditions (top panel). By $t=3.0$ Myr (middle panel), individual sub-clusters undergo stellar collisional runaways with objects more massive than $\msol{300}$ marked using '+' symbols. By $t=7.0$ Myr, most of the sub-clusters have merged into the final central star cluster, while the formed IMBHs ('+' symbols) sink into the centre of the cluster.}
    \label{fig0}
\end{figure}

\subsection{Initial star cluster models}

We outline the key physical properties of the star cluster initial conditions of the simulations of \cite{Rantala2026}. Nearly all (30/33) initial star cluster models used for this study are hierarchical cluster assembly regions that consist of hundreds of individual sub-clusters that merge within a few dynamical times ($\lesssim10$--$20$ Myr) to form a single star cluster. We use three isolated star cluster models to explore the effects of alternative stellar wind models. The basic building block of our hierarchical star cluster assembly regions is a sub-cluster. We use similar spherically symmetric and isotropic \cite{Plummer1911}\footnote{We show in \cite{Rantala2025b} that the Plummer results correspond to \cite{King1962} model outcomes for clusters of similar densities.} models for the sub-clusters and isolated star clusters in our study. The sub-clusters have initial masses up to $M_\mathrm{cl} = \msol{2.5\times10^5}$.

As listed in Table \ref{table: 1}, we explore models with nine different ($N_\mathrm{density}=9$) initial normalisations $f_\mathrm{h}$ of the sub-cluster mass-radius relation \citep{Rantala2024b} while retaining the power-law form $r_\mathrm{h} \propto f_\mathrm{h} M_\mathrm{cl}^\mathrm{\beta}$ of \cite{Brown2021}.
With such normalisations of the mass-radius relation, the sub-cluster population spans a range of $0.08$ pc $\leq r_\mathrm{h} \leq 2.16$ pc in half-mass radii, $\rhosol{3.0\times10^3} \leq \rho_\mathrm{h} \leq \rhosol{6.5\times10^7}$ in 3D half mass density and $\Sigmasol{5.0\times10^3} \leq \Sigma_\mathrm{h} \leq \Sigmasol{3.9\times10^6}$ in half mass surface density. The most massive sub-clusters are the densest among the sub-cluster population, their properties shown in Table \ref{tab: cluster-densities}. The initial cluster densities exceed the observed densities of young massive clusters in the local Universe (e.g. \citealt{Krumholz2019}) and are comparable with the inferred high birth densities of the JWST $z\sim10$ proto globular clusters \citep{Adamo2024}.

For the hierarchical setups the total region mass is $M_\mathrm{region}=\msol{10^6}$ with $N_\star \sim 1.8\times10^6$, and the individual sub-clusters follow the universal power-law $-2$ mass function \citep{Elmegreen1996,Zhang1999,Adamo2020,Claeyssens2026}. The assembly region is modelled after the hydrodynamical star-burst simulations of \cite{Lahen2020, Lahen2025b}, has an initial diameter of $\lesssim100$ pc and collapses on a timescale of the order of $\sim10$ Myr, and more than $50\%$ of the initial mass of the clustered region eventually end up in the central regions of the assembled clusters. For the further structural and kinematic properties of the hierarchical cluster assembly setups see \cite{Rantala2024b}. We use two stellar metallicities: $Z=\zsol{0.10}$ (27 models) and $Z=\zsol{0.01}$ (six models), and perform three random realisations of each simulation setup ($N_\mathrm{random}=3$).

The stars in the individual clusters initially follow the \cite{Kroupa2001} initial mass function (IMF) from $\msol{0.08}$ to a cluster mass dependent IMF cut-off \citep{Weidner2006, Yan2023} which is $\msol{150}$ for massive clusters \citep{Rantala2025b}. The initial binary star population properties are motivated by \cite{Moe2017,Winters2019,Offner2023} and the initial total binary fraction is $f_\mathrm{b}\sim0.29$ with a majority of massive stars in binary systems.

\subsection{EMS/SMS/IMBH formation in dense assembling star clusters}

We simulate the evolution of our hierarchical and isolated star cluster models until $t=7.5$ Myr at which point all collisionally grown stars have reached the ends of their lives. We illustrate the process of hierarchical cluster assembly in Fig. \ref{fig0}. The collapse of the hierarchical region is rapid, only a few dynamical times of the model. By the end of the simulation, most of the sub-clusters have merged into the growing central star clusters, and the remaining sub-structure is erased by $10$--$20$ Myr \citep{Rantala2024b} by tidal effects and dynamical heating from two-body relaxation.

The initial density of the sub-clusters determines if stellar collisions occur in the models. In low-density models HD1--3 with half-mass surface densities below $\Sigma_\mathrm{h} \lesssim \Sigmasol{4\times10^4}$ and 3D densities $\rho_\mathrm{h}\lesssim \rhosol{6\times10^4}$, few collisions occur. In these models, the highest-mass BHs of $m_\bullet \lesssim \msol{60}$ originate from stellar evolution. In initially denser models, increasingly runaway stellar collision cascades occur, the highest IMBH masses depending on the initial cluster properties \citep{Rantala2025b,Rantala2026}. The highest reached stellar masses are in the SMS range ($\gtrsim \msol{10^4}$) in the setup ID9Z1 that has weaker stellar winds, and in the EMS range in the models HD6--9 with fiducial \cite{Vink2018} winds. Immediately after their assembly, massive star clusters of the models HD4--9 host one to a few IMBHs, and a population of stellar BHs.

\section{Results}\label{section: 3}

\subsection{Classification of BH star interactions}

\begin{table*}
    \centering
    \begin{tabular}{llccccc}
        \hline
         Label & Interaction & BH & $m_\bullet$ $[M_\odot]$ & $m_\star$ $[M_\odot]$ & mass ratio & likely outcome \\
         \hline
         A & IMBH+MS/VMS & IMBH & $154$--$4347$ & $9$--$272$ & $m_\star/m_\bullet<1 $ or $m_\star/m_\bullet \ll 1$  & TDE \\
         \hline
         B & IMBH+VMS/EMS & IMBH & $250$--$2716$ & $584$--$5078$ &  $1 \lesssim m_\star/m_\bullet \lesssim 10$ & star disrupted, BH accretion?\\
         \hline
         C & BH+MS & stellar & $14$--$71$ & $10$--$66$ & $m_\star/m_\bullet \sim 1$ & binary interaction or micro TDE\\
         D & BH+VMS & stellar & $19$--$56$ & $31$--$337$ & $1 \lesssim m_\star/m_\bullet \lesssim 10$ & star disrupted, BH accretion?\\
         \hline
         E & BH+EMS/SMS & stellar & $40$--$61$ & $1010$--$24493$ & $22 \leq m_\star/m_\bullet \leq 407$ & BH embedded within star?\\
         \hline
    \end{tabular}
    \caption{The five BH star interaction categories identified from our simulation sample. The columns list the BH and stellar masses $m_\bullet$, $m_\star$ and the mass ratio $m_\star/m_\bullet$ as well as the likely outcome of the interaction.}
    \label{table: 2}
\end{table*}

\begin{figure*}
    \centering
    \includegraphics[width=0.6\textwidth]{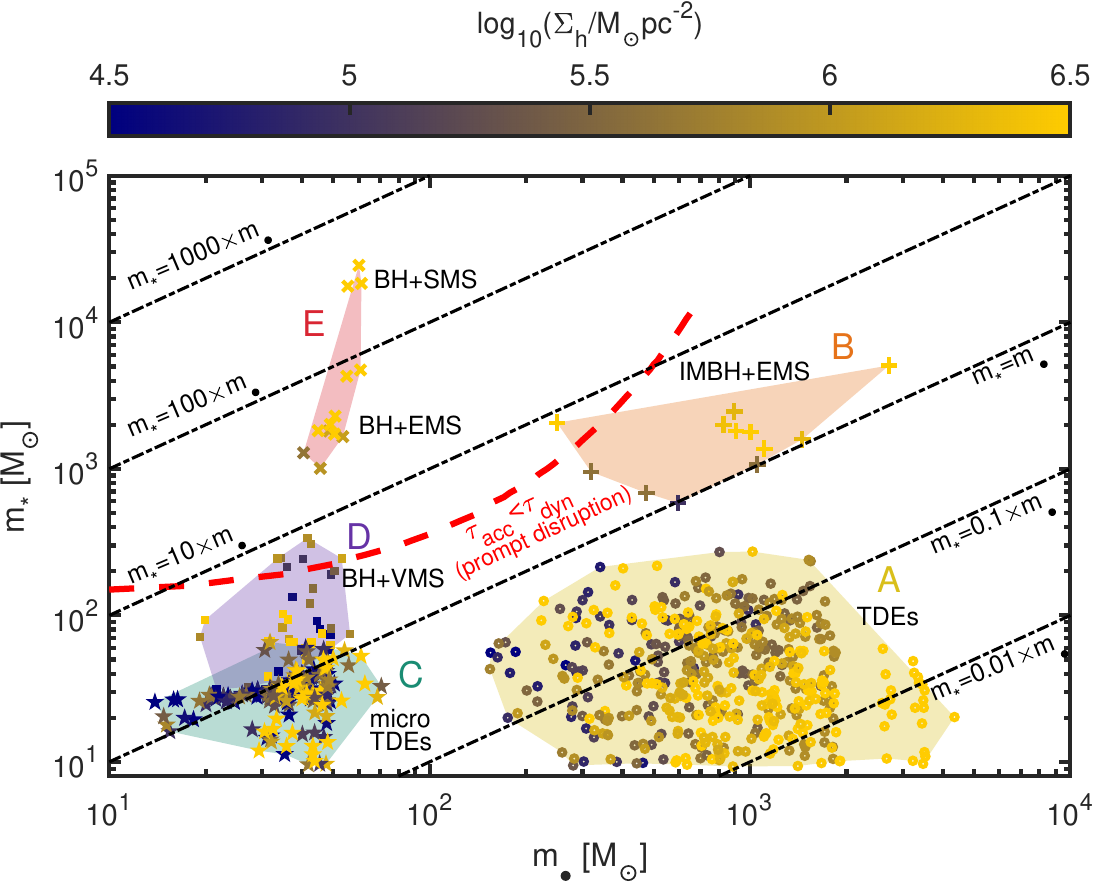}
    \caption{The landscape of strong interactions between massive ($m_\star > \msol{10}$) stars and BHs in the simulation sample. The five interaction categories A--E are largely distinct from each other with only categories C and D somewhat overlapping. The category E (BH+EMS/SMS) stands out from the rest of the interactions by its high stellar masses ($\msol{1010} \lesssim m_\star \lesssim \msol{24493}$), stellar BH masses ($\msol{40}\lesssim m_\bullet \lesssim \msol{61}$) and consequently very high star BH mass ratios of $22 \lesssim m_\star / m_\bullet \lesssim 407$. The recent estimate for prompt disruption in a strong star BH interaction ($\tau_\mathrm{acc} < \tau_\mathrm{dyn}$) by \citet{Baumgarte2025} is displayed as the dashed line. Below the line, the star is very rapidly disrupted and accreted.}
    \label{fig1}
\end{figure*}

Before the end of their lives, ordinary stars, EMSs and SMSs may strongly interact with BHs. Interactions of stars with stellar BHs are possible in all models while an interaction with an IMBH requires a sufficiently dense initial setup (HD4, or denser) in which an IMBH can form. Similarly, interactions involving EMSs/SMSs are not possible in low density cluster models. We present the masses of the stars ($m_\star$) and BHs ($m_\bullet$) that participated in strong BH star interactions in our simulation sample in Fig. \ref{fig1}. Examining the masses, evolutionary phases and remnant types of the stars, we identify in total five largely distinct categories of BH star interactions. These five interaction categories A--E are listed in Table \ref{table: 2}, and elaborated in detail as follows.

\subsubsection{Category A: IMBHs interacting with (very) massive stars}
Most simulation models that form IMBHs (HD4Z2 and denser setups) involve later tidal disruption event driven IMBH growth. In the simulation sample of this study, most interactions between stars and BHs belong to a category that more or less corresponds to classic TDEs: stars on almost parabolic orbits ($e\sim1$) being disrupted by considerably more massive BHs ($m_\bullet \gg m_\mathrm{\star}$). The masses of the IMBHs are in the range of $\msol{154} \lesssim m_\bullet \lesssim \msol{4347}$ while the disrupted stars ($\msol{9} \lesssim m_{\star} \lesssim \msol{272}$) are massive main sequence stars (MS) or VMSs. A small subset ($<4\%$) of the stars have grown by stellar collisions above the initial IMF upper limit of $\msol{150}$. The rate for the IMBH TDEs reaches $\Gamma_\mathrm{TDE}\sim2$--$\collratesol{3\times10^{-5}}$ per assembling cluster in the densest models \citep{Rantala2026}. Fig. \ref{fig1} also shows that TDEs occur more frequently in denser systems, as expected from simple cross section collision rate arguments (e.g. \citealt{Binney2008}).

\subsubsection{Category B: IMBHs interacting with very or extremely massive stars}
Next, we consider stellar disruptions that involve an IMBH and a VMS/EMS. As IMBHs can only form in models denser than HD3Z2 and EMSs in denser than HD5Z2, the category B interactions are absent in low density star cluster setups. In the simulation sample, IMBHs in this interaction category have masses between $\msol{250}\lesssim m_\bullet \lesssim \msol{2716}$ and VMS/EMS masses in the range of $\msol{584}\lesssim m_\star \lesssim \msol{5078}$. The defining difference between interaction categories A and B is the mass ratio of the interaction. While in the category A (IMBH+MS/VMS) typically $m_\bullet \gg m_\star$, in the category B (IMBH+VMS/EMS) $1 \lesssim m_\star/m_\bullet \lesssim10$. This is intrinsically related to the star cluster formation mechanism. As typically only one star can considerably grow by collisions in isolated, idealised clusters \citep{Baumgardt2011,Fujii2013,Rantala2025a}, the IMBH VMS/EMS disruptions of comparable mass ratios are possible only in the hierarchical assembly setup. Moreover, interactions involving high mass EMSs can occur exclusively in the densest systems ($\Sigma_\mathrm{h} \gtrsim \Sigmasol{3\times10^5}$) in which EMS formation is possible \citep{Rantala2026}.

The fact that the EMS is up to ten times more massive than the IMBH signifies that the outcome of the interaction does likely not correspond to a common TDE. Still, we expect that the EMS would be strongly perturbed or disrupted by the IMBH interaction, even though the IMBH may not in reality accrete all of the available disrupted stellar material. We estimate that the rate of the IMBH EMS disruptions is at least $\sim40$ times lower than the TDE rate, below $\Gamma_\mathrm{IMBH-EMS}\lesssim\collratesol{5\times10^{-7}}$ per assembling dense cluster.

\subsubsection{Category C: stellar BHs interacting with (massive) stars}
Next, we consider strong interactions that involve stellar mass BHs ($m_\bullet \lesssim \msol{71}$) and massive stars ($\msol{10} \lesssim m_\star \lesssim \msol{66}$). With $m_\star/m_\bullet \lesssim 2.1$, the category corresponds to so called micro TDEs (e.g. \citealt{Perets2016, Kremer2019, Rastello2026}) with relatively equal mass ratios ($m_\star/m_\bullet \sim 1$). These interactions can occur in all our setups as they do not involve collisionally grown objects. In our simulations most of the category C micro TDE BH star interactions originate between dynamical pairs, as opposed to primordial binary companions, at the centres of the densest star clusters.

\subsubsection{Category D: stellar BHs interacting with very massive stars}

Although the BH star interaction category overlaps with the category C (micro TDEs) in the $m_\bullet$--$m_\star$ parameter space, we distinguish the two categories. In category D, stellar BHs in the mass range of $\msol{19} \lesssim m_\bullet \lesssim \msol{56}$ interact with massive stars and VMSs up to $m_\star \sim \msol{337}$ with mass ratios up to $m_\star/m_\bullet \lesssim 8.1$, considerably larger than for the category C micro TDEs. VMSs with masses beyond $\gtrsim \msol{150}$ need to be built up by stellar collisions, so they do not occur in low density cluster setups. If the VMS is disrupted by the interaction and a considerable fraction ($\gtrsim50\%$) of the VMSs is accreted into the stellar BH, such strong BH VMS interactions provide a channel for the formation of low mass IMBHs (e.g. \citealt{Rantala2024b, Mestichelli2026}).

\subsubsection{Category E: stellar BHs with extremely massive and supermassive stars}
In Fig. \ref{fig1}, the final fifth category E of the BH star interaction classification stands out from the other four interaction event classes. Most importantly, the stellar masses are large, well in the EMS/SMS regime ($\msol{1010} \lesssim m_\star \lesssim \msol{24493}$) while the BH masses are in the stellar remnant range ($\msol{40}\lesssim m_\bullet \lesssim \msol{61}$), resulting in unprecedentedly high interaction mass ratios of $22 \lesssim m_\star / m_\bullet \lesssim 407$ compared to the previously detailed categories. More than $10$ such strong BH EMS/SMS interactions occur in the simulation sample. These all occur in distinct sub-clusters, as in a collision cascade only one star per isolated cluster can considerably grow \citep{Baumgardt2011, Fujii2013, Rantala2024b}. As the channel requires an EMS/SMS, it can only operate in star clusters with high stellar densities above $\Sigma_\mathrm{h} \gtrsim \Sigmasol{3\times10^5}$ (models HD6--9 and ID9Z1). In our densest models HD9 and ID9Z1 with $\Sigma_\mathrm{h}\sim2$--$\Sigmasol{4\times10^6}$, most EMSs (and SMSs in the setups with weaker winds) interact with a stellar BH during their lifetimes. This roughly corresponds to BH EMS/SMS interaction rates of less than $\Gamma_\mathrm{BH-EMS} \lesssim \collratesol{2\times10^{-7}}$ per assembling very dense ($\Sigma_\mathrm{h} \gtrsim \Sigmasol{10^6}$) star cluster.

\begin{figure}
    \centering
    \includegraphics[width=\linewidth]{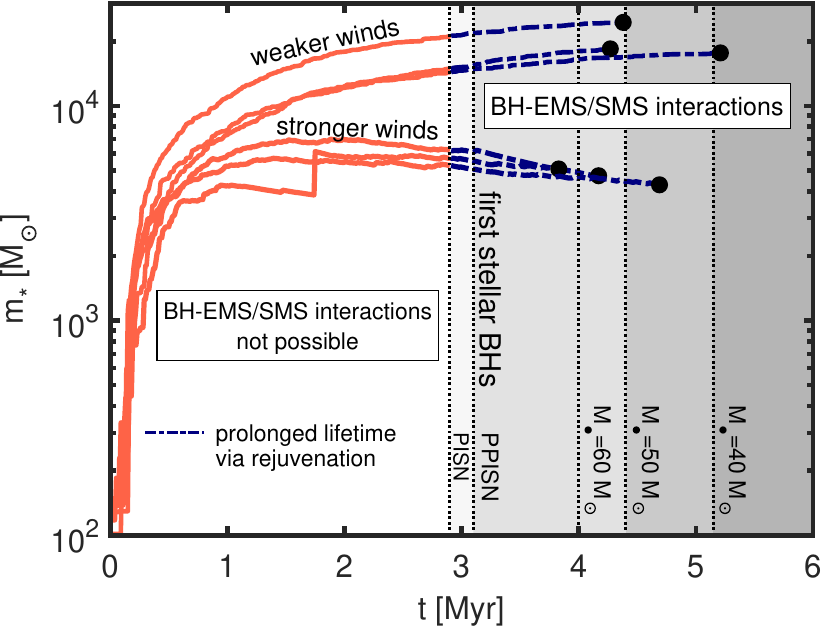}
    \caption{The growth (solid lines) of six EMSs/SMSs in the low metallicity ($Z=\zsol{0.01}$) models HD9Z1 (stronger winds) and ID9Z1 (weaker winds). The vertical lines and the shaded regions indicate the formation times of stellar BHs ($\gtrsim3.1$ Myr) according to their masses. The EMSs/SMSs have their lifetimes extended beyond $\sim3$ Myr via collisional rejuvenation (dot-dashed lines) until they interact with a stellar BH (five category E cases) or an IMBH (one category B case) between $t=3.7$ Myr and $t=5.2$ Myr.}
    \label{fig2}
\end{figure}

\begin{figure*}
    \centering
    \includegraphics[width=\textwidth]{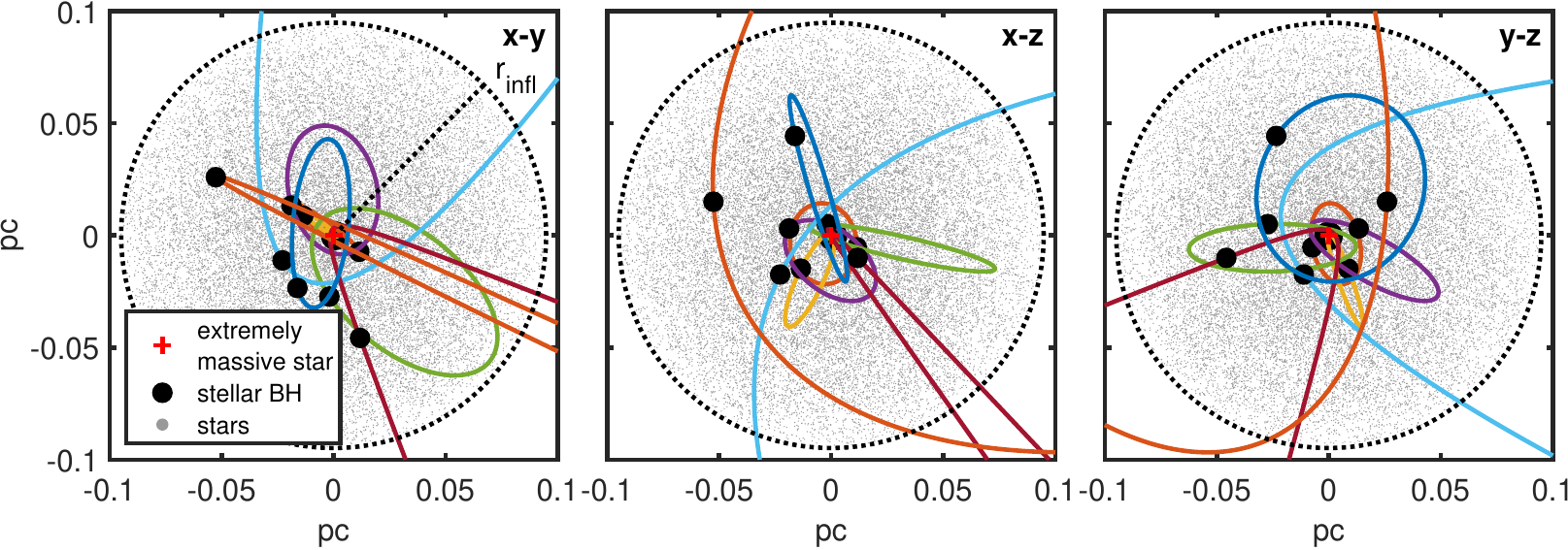}
    \caption{The osculating near-Keplerian orbits (solid lines) of $\sim10$ stellar BHs (filled circles) within the radius of influence of the supermassive star (the central $+$ symbol) at $t=4.35$ Myr. The three panels show x-y, x-z and y-z projections of the SMS sphere of influence (dashed outer circle) while the stars within sphere of influence are represented as small background dots. Note that only the stars within the influence radius are displayed.}
    \label{fig3}
\end{figure*}

\subsection{The time window for BH EMS/SMS interactions}

In Fig. \ref{fig2}, we present the collisional growth histories of six EMSs/SMSs that interacted either with an IMBH (category B; one interaction) or with a stellar BH (category E; five interactions) in the low-metallicity ($Z=\zsol{0.01}$) simulations HD9Z1 (\citealt{Vink2018} winds) and ID9Z1 (weaker winds). The weaker wind models at $t<4$ Myr were presented in \cite{Rantala2026} as a wind prescription comparison to the models of \cite{Vergara2025}, but here we display the full evolution of the collisionally growing stars beyond $t\geq4$ Myr until their first interaction with a BH of any mass.

First massive stars end their lives in pair-instability supernova (PISN) events at $2.9$ Myr $\lesssim t \lesssim 3.1$ Myr and leave behind no compact remnant. First low mass stellar BHs originate at $t\sim3.1$ Myr from pulsational PISN (PPISN) explosions while ordinary core collapse supernovae begin to produce $\msol{60}$ BHs at $\sim4.0$ Myr. Lower mass BHs form from less massive SN progenitors at later times. Meanwhile, the lifetime of the collisionally grown extremely massive stars is always $\lesssim6$ Myr in \cite{Rantala2026}. Thus, we estimate that EMS/SMS BH interactions are possible during a brief time window of only $\lesssim2$ Myr in duration. The EMSs/SMSs of Fig. \ref{fig2} interact with stellar BHs between $3.7$ Myr $\lesssim t \lesssim 5.2$ Myr in a brief time window of $\sim1.5$ Myr.

Stellar BHs are abundant within the sphere of influence of the EMSs at $t\gtrsim3$--$4$ Myr. In our densest star clusters, the EMSs are surrounded by the order of $N_\mathrm{\bullet}\sim10$ by $t\sim4.0$--$4.5$ Myr and $N_\mathrm{\bullet}\sim50$ at $t>5$ Myr. For comparison, the influence radius $r_\mathrm{infl}$\footnote{We define the influence radius as $M_\star(<r_\mathrm{infl})=m_\mathrm{EMS}$ using the cumulative radial mass profile $M_\star(<r)$ centred at the extremely massive star.} of EMSs/SMSs always encloses $\geq10^3$--$10^4$ stars. In Fig. \ref{fig3}, we show the ten stellar BHs and their osculating near-Keplerian orbits within the sphere of influence ($r_\mathrm{infl}\sim0.095$ pc) of the $m_\star \sim \msol{25000}$ SMS from model ID9Z1 at $t=4.35$ Myr. Soon after this, the most tightly bound stellar BH ($a\sim10^{-3}$ pc, $e\sim0.997$) collides with the central SMS. Eight of the ten BHs orbit deep within the sphere of influence and are gravitationally bound to the SMS. The other two BHs on wider orbits are bound to the cluster core including the SMS and its surrounding inner stars.

\subsection{The dynamics of stellar BHs around the EMSs/SMSs}

\begin{figure}
    \centering
    \includegraphics[width=\linewidth]{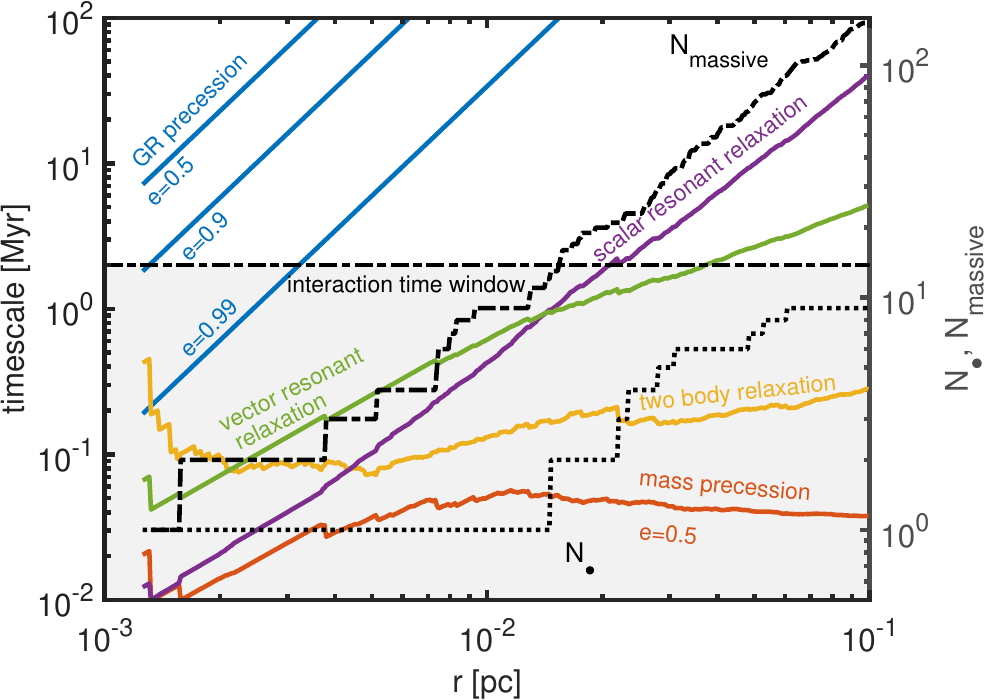}
    \caption{The stellar dynamical precession and relaxation timescales within the influence radius ($r_\mathrm{infl}\sim0.1$ pc) of a $\sim\msol{25000}$ star from the simulation ID9Z1. $N_\bullet\sim10$ BHs and $\sim200$ massive stars orbit within the influence radius ($r_\mathrm{infl}\sim0.1$ pc) of the SMS. Scalar resonant relaxation and non-resonant two-body relaxation can drive stellar BHs into interaction orbits with the SMS on a timescale that is shorter than the interaction time window of $\lesssim2$ Myr.}
    \label{fig4}
\end{figure}

We proceed to identify the physical mechanisms responsible for depositing stellar BHs on collision orbits with the growing EMSs/SMSs. Due to the very rapid ($\lesssim 0.5$ Myr) mass segregation of the clusters, stars with masses beyond $5$--$\msol{10}$ segregate into the cluster centres, and are clearly over-represented in the mass function of the stars colliding with the EMS \citep{Rantala2024b}. The crossing time of the densest clusters is only of the order of $t_\mathrm{cross}\sim0.01$ Myr. Thus, it is very unlikely that a massive star on a loss cone orbit ends its life and produces a loss cone orbit BH, rather, stellar BHs that form outside the loss cone are deposited into it by relaxation processes. We note that the same dynamical mechanisms that place stars on collision orbits with the EMS/SMS act on stellar BHs as well. In our simulations, the orbit of a typical stellar BH interacting with the EMS is wide ($a\sim1\;$mpc$\;\sim200$ AU), and almost parabolic ($e\sim1$), $70\%$ of the BH EMS interactions occur from bound orbits while the rest of the collision orbits are mildly hyperbolic and bound to the cluster. As the time window for BH EMS interactions is only $\lesssim2$ Myr, any relaxation process responsible for filling the EMS loss cone with even a single stellar BH should operate on a $\sim$ Myr timescale.

Fig. \ref{fig4} elaborates the stellar dynamical environment around our most massive SMS ($m_\star \sim \msol{25000}$, model ID9Z1 with weaker winds) at $t=4.35$ Myr just before the strong BH SMS interaction. In the figure, we indicate the dominating orbit precession and relaxation timescales at different separations from the EMS as well as the numbers of stellar BHs and $>\msol{20}$ BH progenitor stars that have not yet ended their lives. The spatial extent of the figure ($0.1$ pc) coincides with the gravitational sphere of influence of the SMS ($r_\mathrm{infl}\sim0.095$ pc). In total $N_\bullet \sim 10$ BHs and $\sim200$ massive stars orbit close to the EMS. Compared to the spatial scale of the influence radius, the radii of the most massive EMSs in our simulations are small \citep{Rantala2026}, of the order of few hundred $R_\mathrm{\odot}$ $\lesssim 10^{-5}$ pc. The various timescale estimates for the precession (relativistic precession $t_\mathrm{GR}$, mass precession $t_\mathrm{M}$) and relaxation (non-resonant two-body $t_\mathrm{NR}$, scalar resonant $t_\mathrm{SRR}$, vector resonant $t_\mathrm{VRR}$) processes are summarized in Appendix \ref{section: appendix1}.

Within the $r_\mathrm{infl}$, mass precession originating from the extended stellar component is the dominant orbit precession mechanism at separations at which the stars and stellar BHs orbit the SMS. General relativistic (GR) precession timescale $t_\mathrm{GR}$ is long at typical separations of innermost stars and stellar BHs around the EMSs/SMSs even for highly eccentric orbits. The mass precession timescale $t_\mathrm{M}$ is always shorter than the GR precession timescale ($t_\mathrm{M} \lesssim 0.1$ Myr for $e=0.5$). We note that the mass precession timescale changes very little with the separation from the central SMS. This is due to the cuspy density distribution within the influence radii of massive objects \citep{Bahcall1976,Bahcall1977,Rizzuto2023}: for a power-law cusp of $\gamma=3/2$ the mass precession timescale is constant \citep{Madigan2009, Merritt2011}. Within the innermost regions ($\lesssim 5\times10^{-3}$ pc) around the EMS the BH orbits are strongly affected by scalar resonant relaxation (SRR). In this region, $t_\mathrm{SRR} \lesssim 0.1$ Myr, and the coherent torques from other orbiting stars and stellar BHs efficiently alter the magnitude of the angular momentum vector of the BHs while keeping their semi-major axes relatively unaffected. The process changes the orbital eccentricities, and thus the pericentre separations of the BH orbits, allowing for strong interactions between stellar BHs and the EMS/SMS. Outside $r\gtrsim5\times10^\mathrm{-3}$ pc, the dominant relaxation mechanism is the non-resonant two-body relaxation ($t_\mathrm{NR} \lesssim 0.5$ Myr) that affects both the semi-major axes and eccentricities, and thus the pericentre separations, in a random walk like manner. Vector resonant relaxation is also rapid ($t_\mathrm{VRR} \lesssim$ a few Myr) within the influence radius of the EMS, but it alters the orientations of the orbits around the EMS and not their pericentre distances.

It is well established that the growth of a central point-like mass such as an SMBH can strongly alter the orbits of the surrounding objects (e.g. \citealt{Gondolo1999,Merritt2013}). We now confirm that this effect is negligible for the BH orbit evolution in our setup. During the interaction window at $t\gtrsim3$ Myr, the EMS/SMS may still grow by repeated collisions from stars within its sphere of influence especially in the models with weaker winds. However, as $m_\mathrm{EMS}/\dot{m}_\mathrm{EMS} \gg t_\mathrm{cross}$, this growth is adiabatic, so the radial action $J_\mathrm{r}$ of the orbiting stellar BHs is conserved, and in addition the setup remains spherically symmetric. The resulting change in the orbital elements of the orbiting BH is $\propto \Delta m_\mathrm{EMS}/m_\mathrm{EMS}$ in which $\Delta m_\mathrm{EMS}$ represents the total mass of the stars accreted from the extended component into the EMS/SMS. We verify the adiabatic estimates using \mstar{} \citep{Rantala2020} few-body simulations including a central SMS ($m_\star = \msol{10^4}$), a power-law extended potential with $M_\mathrm{ext}=\msol{10^3}$ within the orbit ($a=10^{-3}$ pc, $e=0.90$) of a $m_\bullet=\msol{10}$ stellar BH. In the test simulations the SMS completely drains the extended background potential in $1000$ orbital timescales of the BH, conserving the total mass of the system. As the extended potential is drained, the perturbations on the BH orbit become gradually weaker at the apoapsis and stronger at the periapsis as the potential becomes more point-like, adiabatically contracting the orbit. For the modest $\sim10\%$ SMS mass growth during the interaction window, as in Fig. \ref{fig2}, the semi-major axis of an orbiting BH mildly shrinks ($\Delta a/a \sim -0.1\%$) and the orbit becomes mildly more eccentric ($\Delta e \sim +0.01$). Nevertheless, both the scalar resonant relaxation and non-resonant two-body relaxation have considerably stronger effect on the BH orbit compared to the adiabatic SMS growth.

\subsection{The nature of the BH EMS/SMS interactions}

Strong interactions of BHs and stars with comparable masses have been well studied in two regimes: micro TDEs \citep{Perets2016, Fragione2019, Kremer2019, Wang_Yi_Han2021b, Kremer2022, Ryu2022, Ryu2023, Vynatheya2024, Xin2024, Rastello2026, Tsuna2025} linked to high energy and optical transients, as well as common envelope interactions involving a BH (e.g. \citealt{MacLeod2015, Oncins2022}) that may lead to the formation of compact X-ray binaries (e.g. \citealt{Ivanova2005b, Ivanova2017}). Based on the large mass ratios of the SMS BH interactions up to $m_\star/m_\bullet\gtrsim400$, we expect that the process is more comparable to a peculiar tidal capture \citep{Press1977} followed by an inspiral rather than a tidal disruption event. Hydrodynamical simulations of Thorne-\.{Z}ytkow formation via the collision channel \citep{Williams2025} support this picture: collisions of NSs with stars of $\msol{5}$ and above lead to stable configurations of stars with NS cores. Approaching from a marginally bound, near parabolic (zero energy) orbit, the stellar BH excites tides and oscillations in the SMS, losing orbital energy which the star can radiate away, and eventually sink into the gaseous atmosphere of the star. Then, due to gas drag forces, the BH will inspiral and settle into the core of the SMS. 

We show the recent estimate of \cite{Baumgarte2025} for the outcome of collisions between BHs and very massive stars as red dashed line in Fig. \ref{fig1}. At the line, the accretion timescale $\tau_\mathrm{acc}$ equals the dynamical timescale of the star $\tau_\mathrm{dyn}$. Below the line, $\tau_\mathrm{acc} < \tau_\mathrm{dyn}$, and prompt stellar disruption is expected. This essentially covers four of our five categories, A--D. The category E lies $\sim1$--$2$ orders of magnitude (in stellar mass) above the prompt disruption curve. As such, we suggest that the likely outcome of an SMS BH interaction in category E is not the prompt destruction of the star, but at least a somewhat longer-lived configuration. However, we note that even if the massive star survives, it likely loses mass depending on its structure \citep{RamirezGaleano2025,RomanGarza2026} while acquiring its stellar BH core. The initially $\lesssim \msol{60}$ BH core is expected to grow rapidly by accretion \citep{Hassan2026}, and consume the stellar envelope in $\lesssim1$--$2$ Salpeter times ($t_\mathrm{Salp}\sim45$ Myr). We note that this $\sim100$ Myr timescale \citep{RomanGarza2026b} is long enough for accreting subsequent stellar BHs into the QS. Thus, we expect that the star cluster $\rightarrow$ SMS $\rightarrow $ QS pathway allows a class of gas embedded gravitational wave mergers.

\section{Discussion: the formation and the early evolution of faint multiple LRD systems}\label{section: 4}

Observationally, the expected bolometric luminosity of a $\sim\msol{10^4}$ SMS or QS is of the order of its Eddington luminosity $L_\mathrm{QS} \sim 3\times10^8 \Lsun$ \citep{Santarelli2026,RomanGarza2026b} corresponding to $\log_\mathrm{10}(L_\mathrm{QS}/\textrm{erg\:s$^{-1}$})\sim42.1$. As such, the QS like SMSs in our simulations with masses up to $\msol{2.5\times10^4}$ with luminosities $\lesssim 8\times10^8 \Lsun$, or $\log_\mathrm{10}(L_\mathrm{QS}/\textrm{erg\:s$^{-1}$})\sim42.5$, are less luminous than typical LRDs with $\log_\mathrm{10}(L/\textrm{erg\:s$^{-1}$})\sim44$ ($L \sim 3\times10^{10} \Lsun$) (e.g. \citealt{Akins2025, Greene2026, Umeda2026}). To explain the luminosity of a typical LRD, a QS mass of $m_\mathrm{QS}\sim \msol{10^6}$ would be required. However, our inferred QS luminosities are consistent with luminosities of the potential emerging population of faint LRDs \citep{Yanagisawa2026} that can be detected in strongly magnified star forming galaxies at high redshift.

\begin{figure}
    \centering
    \includegraphics[width=\linewidth]{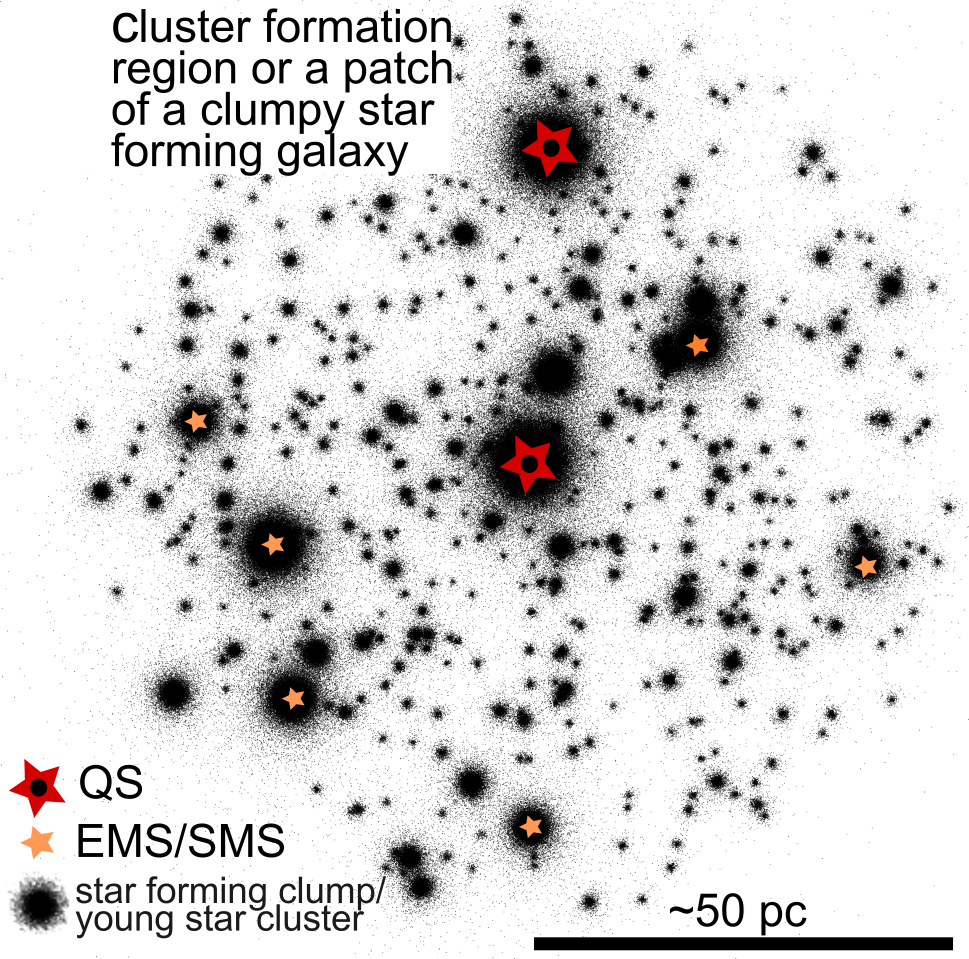}
    \caption{A schematic illustration interpreting a hierarchically assembling star formation region as a dual (or a multiple) system of faint LRDs. The densest and most massive sub-clusters form extremely massive or supermassive stars (EMS/SMS) via runaway stellar collisions, and a subset of these acquire stellar BH cores, becoming long-lived quasi-stars (QSs) responsible for the luminosity of the faint LRD pair. The young, dense star clusters or star forming clumps within the region that do not contain SMSs/QSs appear as blue companions to the faint LRDs.}
    \label{fig5}
\end{figure}

In Fig. \ref{fig5}, we interpret a faint dual LRD system similar to the "Red Eyes" of \cite{Yanagisawa2026} as an early ($t\lesssim10$ Myr) evolutionary phase of a clustered hierarchically assembling star formation region. Our clustered region in its early evolution shares remarkable similarities to the Red Eyes system. First, the Red Eyes LRD pair separation of $\sim70$ pc and its off-centre nature is consistent with the spatial sizes and locations of star cluster assembly regions \citep{Lahen2019,Lahen2020,Rantala2024b,Lahen2025b}. \cite{Yanagisawa2026} estimate that the faint LRD pair will merge rapidly, consistent with the swift assembly of the star formation regions into single clusters. Second, the bolometric luminosities of the faint Red Eyes LRDs are $3.3\times10^{42}$ erg$\:\mathrm{s}^{-1}$ and $6.8\times10^{42}$ erg$\:\mathrm{s}^{-1}$ corresponding to inferred BH masses of $1.3$--$\msol{2.5\times10^4}$ \citep{Yanagisawa2026}. At their early evolutionary phase, our QSs have masses up to $\sim\msol{2.5\times10^4}$, with estimated maximum luminosities of $\log_\mathrm{10}(L/\textrm{erg\:s$^{-1}$})\sim42.5$. Third, the environment of the faint LRD pair is consistent with a clustered star forming region, or a patch of a clumpy star forming galaxy. The compact host galaxy of the Red Eyes system has a total stellar mass of $\sim \msol{10^8}$, a stellar population age of only $90$ Myr, and contains multiple star forming clumps. In Fig. \ref{fig5}, the most massive and dense sub-clusters of the hierarchically assembling star formation region form extremely massive or supermassive stars, and a sub-set of the EMS/SMSs acquires a stellar BH core to become long lived QSs. Consequently, in our picture, faint LRDs from hierarchical cluster assembly are expected to be accompanied, besides other faint LRDs, by young, blue, massive star forming clumps and star clusters that do not contain SMSs, QSs, or massive BHs. At $t=5$ Myr, the bolometric luminosity of the stellar population in the cluster assembly region ($\msol{10^6}$ within 100 pc) is, ignoring any attenuation, $L_\mathrm{\star}\sim5\times10^8 \:L_\odot = 2\times10^{42}$ erg$\:\mathrm{s}^{-1}$ \citep{Leitherer1999, Bressan2012, Hawcroft2025}. The stars in the initially most massive individual sub-cluster provide up to $\sim1/4$ of this luminosity. If the star formation in the region ceases, the stellar luminosity of the region rapidly declines, leaving the light of the system QS increasingly dominated. For the estimated spectral energy distribution (SED) of our QS+cluster model, we do not provide an SED model of our own, but refer to existing SEDs in the literature. First, the model A2 of \cite{Martins2020} (the left panel of their figure 10) includes a proto globular cluster containing a cool SMS, a setup very similar to our model, with the cluster and SMS masses within a factor of $\sim2$ of our setup. Most interestingly, the SED of model A2 shows an LRD like spectrum with a V-shape around the Balmer break with the SMS light dominating the SED around the Balmer break, and at redder wavelengths. Second, the scenario of \cite{Inayoshi2026} involving a growing seed SMBH with a coeval young stellar population results in an SED similar to \cite{Martins2020} while evolving on a $\sim15$ Myr timescale reminiscent of our model.

We emphasize that the current models for SMS and QS SEDs struggle to explain the broad emission lines (e.g. \citealt{Hviding2025, Ji2026}) up to $2000$--$3000 \: \textrm{km s}^{-1}$ observed in typical bright LRDs, unless the lines are not broadened by gas kinematics but by radiative transfer effects in the dense gas such as electron scattering (e.g. \citealt{Begelman2026,Rusakov2026,Takasao2026}). However, the electron scattering scenario remains debated \citep{Brazzini2025,Madau2026,Scholtz2026}, which may suggest that models beyond the standard QSs are required for gas embedded BHs if they are responsible for the characteristic LRD emission. We also note that the SMSs/QSs in our models form through runaway stellar collisions and continue to experience repeated collisions in the QS phase. This likely results in observational properties somewhat different from the steady state QSs that originate from direct collapse or Pop-III accretion, but evaluating the full consequences of the stellar collisions for the SED and variability of the QS is beyond the scope of this study.

Interpreting faint dual and multiple LRD systems as assembling hierarchical star formation regions, or dense patches of a clumpy host galaxy, has implications for their evolution and descendants. As the QSs grow in mass by further stellar collisions (or gas accretion) thanks to their extended lifetimes compared to SMSs, their luminosities increase. Meanwhile, as the cluster assembly continues, the clumpiness of the region becomes less prominent (see figure 4 of \citealt{Rantala2024b}). Thus, our hierarchical cluster assembly scenario predicts that faint LRDs should frequently have companions (other faint LRDs, star clusters and star forming clumps) within $\lesssim100$ pc, while evolved LRDs with higher QS/BH masses and higher luminosities should be more commonly found in relative isolation.

\section{Conclusions}\label{section: 5}

We have shown that if extremely massive ($\gtrsim\msol{1000}$) or supermassive stars ($\gtrsim\msol{10000}$) can be built up by runaway stellar collisions in dense star clusters, they are likely to strongly interact or collide with at least one stellar mass black hole during their lifetime. During a brief time window of less than $1$--$2$ Myr, EMSs, SMSs, and stellar BHs can co-exist at the centres of the dense star clusters with $N_\bullet \sim 10$--$50$ BHs within the influence radius of the collisionally built up stars. In the innermost regions within the sphere of influence, the scalar resonant relaxation timescale is very short ($t_\mathrm{SRR} \lesssim0.1$ Myr) and SRR can drive stellar BHs on extremely eccentric low pericentre orbits that may lead to strong interactions with the EMS/SMS. Outside this very central region, non-resonant two-body relaxation is responsible for providing stellar BHs into the loss cone of the EMS/SMS while the effect of the adiabatic growth of the EMS/SMS on the stellar BH orbits is minor.

In the strong interactions, the EMSs/SMSs may be up to $\sim400$ times more massive than the stellar BHs. BHs on bound or mildly hyperbolic orbits would likely lose their orbital energy to the EMS/SMS via tidal excitations and gas drag forces, a process resembling a tidal capture \citep{Press1977}, shrinking the orbit until the BH becomes fully embedded into the envelope of the star. Subsequently, the BH spirals into the EMS/SMS, resulting in mass loss from the envelope of the star \citep{RamirezGaleano2025,RomanGarza2026}. Thus, it is likely that the EMS/SMS is not promptly destroyed in the interaction, but instead it simply loses mass, and acquires a stellar BH core. This represents essentially a quasi-star like hydrostatic configuration in which a BH is embedded within the gaseous layers of the star. After becoming embedded, the stellar BH rapidly increases its mass by accretion \citep{Hassan2026}. We propose that such an evolutionary SMS $\rightarrow$ QS pathway is common in the densest ($\Sigma_\mathrm{h} \gtrsim \Sigmasol{10^6}$) star clusters in which SMSs can form. Meanwhile, in intermediate density clusters, less massive and less rejuvenated EMSs can reach the ends of their lives forming IMBHs before colliding with stellar BHs.

Besides stellar collisions, EMSs and SMSs are expected to grow by gas accretion (e.g. \citealt{Gieles2018, RomanGarza2026b}) which our collisional model does not consider. In such accretion dominated models, the accretion rate determines the evolution and the final fate of the star. At low accretion rates below $\lesssim \ratesol{0.02}$ (e.g. \citealt{Haemmerle2018, RamirezGaleano2025}) the SMS does not necessarily reach the GR instability and develop a QS like BH core while in the main sequence before hydrogen depletion \citep{Herrington2023,RomanGarza2026b}. Our dynamical scenario thus provides an alternative pathway for the formation of QS like main sequence phase SMSs with BH cores at low gas accretion rates.

The embedded BH phase alters the standard EMS/SMS $\rightarrow$ IMBH formation pathway. First, the QSs may be considerably more long-lived than SMSs: a QS likely lives (e.g. \citealt{Santarelli2026, RomanGarza2026b, Hassan2026}) of the order of $1$--$2$ Salpeter times ($t_\mathrm{Salp}\sim45$ Myr), considerably longer compared to lifetimes ($\sim$ a few Myr) of SMSs. The extended lifetime, not modelled in our current simulations, may enable continued growth by stellar collisions towards $\sim \msol{10^5}$ within the host star cluster before the BH consumes the entire QS envelope. If substantial, QS growth by stellar collisions may also extend the QS lifetimes beyond the point at which supernovae quench the gas inflow driven growth (e.g. \citealt{Inayoshi2026}). However, the prolonged lifetime contributes only little to the total elemental enrichment budget of the SMS/QS as the BH core likely suppresses nuclear reactions in the star \citep{RomanGarza2026b}.

The expected bolometric luminosities of our $\sim \msol{10^4}$ QSs are at least an order of magnitude lower than the typical LRDs which have luminosities of the order of $\log_\mathrm{10}(L/\textrm{erg\:s$^{-1}$})\sim44$ \citep{Akins2025,Greene2026,Umeda2026}, however, we note that including gas accretion in our setup, or modelling systems with larger masses or higher initial densities, would lead to larger SMS/QS masses and thus higher luminosities. Still, we find that the properties of our hierarchical star cluster assembly regions (see also \citealt{Dekel2025}) with QSs are consistent with faint multiple LRD systems such as the recently discovered $70$ pc faint dual LRD "Red Eyes" \citep{Yanagisawa2026}. First, the $70$ pc LRD separation closely matches the typical distance of sub-clusters in the hierarchical star cluster assembly regions \citep{Lahen2019, Lahen2020, Rantala2024b, Lahen2025b} in low metallicity starbursts. Second, the inferred QS luminosities are of the order of $\log_\mathrm{10}(L/\textrm{erg\:s$^{-1}$})\sim42.5$, consistent with the Red Eyes system. 

Finally, not all sub-clusters form SMSs or QSs. The clustered environment provides multiple young, massive, blue, potentially star-forming clumps and clusters in the vicinity of the QS powered faint LRDs as the universal cluster mass function follows a power-law with a slope of $-2$ \citep{Elmegreen1996, Zhang1999, Adamo2020, Claeyssens2026}. The hierarchical star cluster assembly picture allows to estimate the future evolution of the multiple LRD systems. We propose that the faint multiple LRD phase represents the initial stages of the LRD evolution, and as the region collapses (see also \citealt{Shi2024}), the evolved more luminous LRDs should show fewer close ($\lesssim100$ pc) bright companions.

Our SMS $\rightarrow$ QS scenario from runaway stellar collisions highlights the interconnectedness of the various proposed SMBH seeding mechanisms in dense low metallicity environments. While the QS model is most commonly associated with the DCBH or accreting Pop-III SMS pathways, we have shown that QSs can also form if the DCBH channel fails through intermediate steps involving dense, clustered star formation, and an SMS. While gas embedded massive BH seeds appear to be a robust prediction of both the DCBH and runaway star cluster channel for SMBH formation, the initially lower QS masses from the runaway channel ($\sim\msol{10^4}$) as well as their number densities and small-scale sub-kpc clustering may provide a pathway to distinguish between the two models. We expect that observations of gravitationally lensed LRDs (e.g. \citealt{Furtak2023, Furtak2024, Merida2025, Golubchik2025, Baggen2025}) will play a particularly powerful role in shedding light on the issue.

\section*{Data availability statement}
The data relevant to this article will be shared on reasonable request to the corresponding author.

\section*{Acknowledgements}
AR thanks Jaime Roman-Garza, Natalia Lahén and Takumi Tanaka for useful discussions during the preparation of the article. This article focuses on a sub-sample of numerical simulations that were originally performed by \cite{Rantala2026} using German facilities hosted by the Max Planck Computing and Data Facility (MPCDF), Garching, and the JUWELS Booster of the Jülich supercomputing centre (GCS project 59949 frost-smbh-origins). AR acknowledges support from the Kavli Foundation.


\bibliographystyle{mnras}
\interlinepenalty=10000
\bibliography{manuscript}

\appendix
\section{Orbit precession and relaxation timescales}\label{section: appendix1}

We briefly summarise the most important precession and relaxation timescales for orbits around massive objects in star clusters. For a more thorough stellar dynamical picture see e.g. \cite{Merritt2011, Merritt2013} and \cite{Rantala2024a}.
\subsection{Precession}
Consider an extremely massive star with mass $m_\mathrm{EMS}$ and radius $R_\mathrm{EMS}$. A stellar BH with mass $m_\bullet \ll m_\mathrm{EMS}$ orbits the EMS within its sphere of influence with a semi-major axis $a_\bullet$ and eccentricity $e_\bullet$ with $r_\mathrm{p} = a_\bullet (1-e_\mathrm{\bullet}) > R_\mathrm{EMS}$. The general relativistic precession timescale for the orbit is
\begin{equation}
    t_\mathrm{GR} = \frac{c^2 a_\bullet (1-e_\bullet^2) P_\bullet}{3 G M_\mathrm{EMS}}
\end{equation}
in which $P_\bullet$ is the orbital period of the BH around the EMS. The precession is famously prograde \citep{Einstein1915}. On the other hand, the extended spherical stellar distribution around the EMS ($M_\star(r) \ll m_\mathrm{EMS}$) induces orbit precession to the retrograde direction. The timescale for this mass precession \citep{Merritt2011} is
\begin{equation}
    t_\mathrm{M} = \frac{m_\mathrm{EMS} P_\bullet}{M_\star(<a_\bullet)} f(e_\bullet) = \frac{m_\mathrm{EMS} P_\bullet}{M_\star(<a_\bullet)} \frac{1+\sqrt{1-e_\bullet^2}}{\sqrt{1-e_\bullet^2}}.
\end{equation}
For a spherical power-law cusp $\rho(r) \propto r^\mathrm{-\gamma}$ the mass precession timescale becomes
\begin{equation}
    t_\mathrm{M} \propto a_\bullet^\mathrm{\gamma-3/2} f(e_\bullet).
\end{equation}
Thus, for inner cusps with $\gamma=3/2$ the mass precession timescale does not depend on the semi-major axis of the orbit.

\subsection{Relaxation}

Non-resonant two-body relaxation changes both the energies and angular momenta (and thus the semi-major axes and eccentricities) of the orbiting BHs in a random walk like manner. The timescale for the two-body relaxation process around an EMS is
\begin{equation}
    t_\mathrm{NR} \sim \frac{1}{\beta^2 N} \frac{m_\mathrm{EMS}^2}{\Tilde{m}^2} P_\bullet
\end{equation}
in which $\Tilde{m} = \langle m^2 \rangle/\langle m \rangle$ is calculated from the stellar mass distribution and $\beta^2=\ln{(m_\mathrm{EMS}/m_\bullet)}$ \citep{Binney2008}. 

In the innermost regions inside the influence radius of the EMS, the number of orbiting bodies is small, and the torques on the BH orbits are correlated and build up in a coherent manner \citep{Rauch1996}. While this resonant relaxation (RR) leaves the energy (semi-major axes) of the orbits relatively unaffected, it can efficiently alter both the direction (vector resonant relaxation; VRR) and the magnitude (scalar resonant relaxation; SRR) of the angular momentum vector. From the point of view of the EMS BH collisions the SRR is the more relevant mechanism as it can efficiently alter the eccentricities and the pericentre distances of the BH orbits. When the mass precession is the dominant orbit precession mechanism, the SRR timescale is approximately \citep{Rauch1996}
\begin{equation}
    t_\mathrm{SRR} \sim \frac{m_\mathrm{EMS}}{\Tilde{m}} P_\bullet.
\end{equation}
Similarly, the VRR timescale can be estimated \citep{Eilon2009, Kocsis2015, Fouvry2019} as
\begin{equation}
    t_\mathrm{VRR} \sim f_\mathrm{VRR} \frac{m_\mathrm{EMS}}{\left[M_\star(a_\bullet) \Tilde{m} \right]^{\mathrm{1/2}}} P_\bullet
\end{equation}
in which the parameter $f_\mathrm{VRR}$ depends on the eccentricity distribution of the orbits and typically gets values in the range of $0.5 \lesssim f_\mathrm{VRR} \lesssim 5$ \citep{Kocsis2015}.


\bsp	
\label{lastpage}
\end{document}